\documentclass[twocolumn,pra,aps,flushbottom,showpacs]{revtex4-1}
\usepackage{xspace,amsmath,amsfonts,amsthm,amssymb,amsbsy,graphicx,color}

\begin{document}

\title{Quantum measurement and the first law of thermodynamics: \\ the energy cost of measurement is the work value of the acquired information}

\author{Kurt Jacobs}

\affiliation{Advanced Science Institute, RIKEN, Wako-shi 351-0198, Japan
\\ Department of Physics, University of Massachusetts at Boston, Boston, MA 02125, USA } 

\begin{abstract}  
The energy cost of measurement is an important fundamental question, and may have profound implications for quantum technologies. In the context of Maxwell's demon, it is often stated that measurement has no minimum energy cost, while information has a work value. However, as we elucidate, the first of these statements does not refer to the cost paid by the measuring device. Here we show that it is only when a measuring device has access to a zero temperature reservoir --- that is, never --- that measurement requires no energy. To obtain a given amount of information, all measuring devices must pay a cost equal to that which a heat engine would pay to obtain the equivalent work value of that information.   
\end{abstract}


\maketitle 

In this paper we wish to determine if, and when, the act of making a quantum measurement requires a minimum energy. It turns out that this question is intimately connected to Maxwell's demon, a machine that is able to extract work from a system in thermal equilibrium by making a measurement upon it~\cite{Plenio01, Maruyama09, Bennett82, Zurek84, Lubkin87, Leff94, Lloyd97, Scully01, Kieu04, Alicki04, Bender05, Quan06, Sagawa08, Jacobs09c, Sagawa09, Rio11, Dahlsten11}. Maxwell realized that the ability to measure the individual velocities of the molecules in a gas in thermal equilibrium would make it possible to bypass the usual laws of thermodynamics, turning the heat of the gas directly into work, and apparently breaking the second law of thermodynamics. This paradox was not fully resolved until 1982, when Bennett~\cite{Bennett82}, building on works by Szilard~\cite{Szilard29} and Landauer~\cite{Landauer61}, showed that the demon must store the measurement results in its memory, and must increase the entropy of the environment when it erases this information~\cite{Landauer61, Shizume95, Piechocinska00, Vedral00, MyEprint, Maroney09, Turgut09}. 

Here we show that measurements have a minimum energy cost, by which we mean a work cost paid by the measuring device. In the following we prove our results rigorously, but the argument can be summarized simply. The demon cannot extract work from a system if its memory, which is also the measuring device, is at the same temperature as the system. If one is concerned with the energy required for measurement, then one cannot simply give the demon a low-temperature reservoir to cool its memory, since creating such a reservoir requires a refrigerator, which requires energy. One must make the measurement without a cold reservoir, and this requires reducing the free energy of the demons memory (the measuring device). The maximal work that can be extracted is exactly the free energy lost by the measuring device. Thus after the work is extracted, the process is energy neutral, and energy has been taken from the measuring device which is the cost paid. If the demon does have a cold reservoir, then the energy cost is less, being the same as that paid by a heat engine to extract work equal to the value of the information. This energy paid can then be replaced if one wishes, and doing so is precisely the erasure part of the usual analysis of Maxwell's demon. As we will explain below, the reason that the cost of measurement is usually stated to be zero is that the usual analysis is not concerned with an energy exchange between the system and the demon, but between these two and the environment. We note that our analysis here is quite distinct from that of Sagawa and Ueda~\cite{Sagawa09} for the same reason. 

Before we proceed, we should be precise about what we mean by ``information''. We define information as knowledge about which pure state a quantum system is in. For this purpose the von Neumann entropy, S, is the appropriate measure of uncertainty, or lack of knowledge, as it gives the entropy that is due purely to classical mixing of a set of basis states, and thus the lack of classical information about the quantum state. Since the maximum von Neumann entropy of an $N$-dimensional system is $\ln N$, we can quantify an observer's information about the system by $I(\rho) = \ln N - S(\rho)$, where $\rho$ is the observer's density matrix. The purpose of a measurement is to increase (on average) the observers information about the system. A measurement has a number of outcomes, and the (average) information provided by the measurement is the difference between the initial and final von Neumann entropy of the system, averaged over the measurement outcomes. 

For readers less familiar with Maxwell's demon, we now review briefly the standard analysis~\cite{Maruyama09, Plenio01}. Consider a container with a partition in the center that divides it into two separate compartments. Both compartments are filled with a gas at equilibrium at the same temperature and pressure. Now enter the demon. The little guy operates a tiny trapdoor in the partition, and constantly looks to see if any gas molecules are heading towards it. If a gas molecule comes towards the trapdoor in the right half of the container, then the demon opens the trapdoor and lets it through to the left. If a molecule comes from the left half, then the demon closes the trapdoor. In this way the demon reduces the volume of the gas, increasing the pressure. Now work can be extracted by allowing the gas to expand to its original volume. For each incoming molecule the demon must store one bit of information, because it has to perform one of two possible actions. By analyzing how a physical system performs such contingent actions, it is possible to see that the demon cannot do so without storing the result of the measurement. However, the need for this storage is very basic --- it is an immediate consequence of the reversibility of the laws of physics. Reversibility requires that all initial states must give a unique final state (otherwise the evolution could not be reversed). In placing all the molecules into half their initial volume, the demon has greatly reduced their available states. But the number of final states must be equal to the number of possible initial states, so the demon's memory needs to make up the difference~\cite{Maruyama09}. Since physical processes are reversible, to erase the demons memory its state must be dumped into the environment. This increases the environment's entropy and thus requires pouring heat into it~\cite{MyEprint}.  

We now analyze the energy dynamics of the measurement process. Our starting point is that, like the ``target'' system on which the demon will act, the demon itself is a physical system and so must begin at thermal equilibrium. Since it is only the demon's memory that we need to consider, we will treat the demon and its memory as synonymous. Let us first consider a demon acting on a target system when both are at the same temperature, $T$. The measurement of the target by the demon consists of an interaction that correlates the demon's states with those of the target. After the interaction, in the subspace defined by each orthogonal pure state of the demon's memory, the target is in a different state. These $N$ states correspond to the $N$ measurement results. The demon can now apply a Hamiltonian in the joint space that performs a different action for each subspace, and thus each final state of target. This is simply the coherent version of measurement combined with feedback control~\cite{Jacobs09c}. It is this conditional action that allows the demon to use the information provided by the measurement to extract work. It turns out that whatever kind of measurement the demon makes, weak or projective, the feedback allows it to extract work equal to $T \Delta I $, where $\Delta I$ is the average reduction of the target's entropy due to the measurement~\cite{Jacobs09c}. 

When the feedback on the target is completed, and the work extracted, the target is back in its initial state (so as to close the cycle)~\cite{Maruyama09, Plenio01}, and thus the memory is no longer correlated with the target. In the appendix we prove that, irrespective of the initial states of the target and memory, the increase in the entropy of the memory is greater than or equal to $\Delta I$. What interests us here is the energy involved in the measurement and feedback, rather than the entropy. But the lower bound on the increase of the memory's entropy allows us to put a lower bound on the increase in its energy. We note that the memory starts in the Boltzmann state, and this state has the special property that it has the maximum entropy given it's average energy. Thus when the entropy of the memory increases, so must it's average energy. The final state of the demon may not be a Boltzmann state. But from the above property of the Boltzmann state, we know that its average energy must be no less than the average energy it would have, if it were in the Boltzmann state with the same entropy. To place a lower bound on the final energy, we therefore need to know the energy difference between the initial Boltzmann state with temperature $T_{\mbox{\scriptsize i}} = T$ and that with the final temperature $T_{\mbox{\scriptsize f}}$. This is given by calculating the energy added by a reversible process that adds entropy $\Delta I$, being~\footnote{Since $E = F + TS$, where $F$ is the Helmholtz free-energy, the change in energy due to a change in temperature (at constant volume) is $dE = dF + S dT + T dS$. By using the explicit expressions for $F$ and $S$ in terms of the partition function, one obtains $dF + S dT = 0$ and thus $dE = T dS$.}
\begin{equation}
     E_{\mbox{\scriptsize min}} = \int_{T_{\mbox{\tiny i}}}^{T_{\mbox{\tiny f}}} \!\!\! T \, dS(T) = \int_{S_{\mbox{\tiny i}}}^{S_{\mbox{\tiny f}}} \!\!\! T(S) \, dS \; \geq \; T_{\mbox{\scriptsize i}} \Delta I  , 
\end{equation}
where the last inequality follows because the entropy increases monotonically with temperature. Thus the heat added to the demon in the measurement and feedback process is no less than $E_{\mbox{\scriptsize min}} \geq T \Delta I$, and so is at least as large as the work that the demon extracts. Note that the energy added is heat, because it is due to an increase in entropy --- no net work need be done on the memory, and any that is added can be subsequently extracted. 

We can conclude that if the memory is in thermal equilibrium at the same temperature as the target, then it cannot extract any work by making a measurement. It is interesting to note that there is a sense in which it can still make the measurement and obtain information, but this information has no work value. Clearly the same argument holds when the memory has a higher temperature than the demon. Does this mean that the demon's memory must be cooled in order to extract work? The answer is no, as the demon has one other available option: it can change its energy levels rapidly so as to take itself out of equilibrium.

Consider what happens if the demon reduces the energies of all its energy levels to that of its ground state. If it does this quickly, it reduces its own average energy, and preserves its entropy. (Note that it must do this quickly, because if it re-thermalizes it will be in the completely mixed state, in which case it cannot extract entropy from the system). If the demon now performs the measurement on the system, because all its energy levels are equal, the increase in its entropy induced by the measurement does not increase its energy. The demon can now extract net work from the system. (But the cycle is not closed until the energy levels of the memory are returned to their initial values.) 

So let us consider a full cycle in which the demon extracts work. Since the demon finishes with all its energy levels in the ground state, we obtain a closed cycle by starting with the demon in this configuration. Since all the energy levels are the same, the entropy of the demon is the maximal value $S_{\mbox{\scriptsize d}}^{max} = \ln N$. To extract work, the demon must reduce the entropy of the target, and the result above shows that it can only do this if its entropy is less than maximal. We can reduce the entropy of the system by increasing one or more of its energy levels above the ground state, and letting it re-thermalize. Since the thermal populations of each level decrease monotonically with energy, we expend the least amount of work by raising the energy levels slowly (quasi-statically) so that the system is always at equilibrium, and the process is thus isothermal. The total amount of work done to raise the energy levels, $W_{\mbox{\scriptsize d}}$, is the difference between the initial and final average energies of the demon, along with the amount of energy poured into the bath. Setting the ground state energy to be zero, we have $W_{\mbox{\scriptsize d}} = \langle E \rangle + T \Delta S$, where $\langle E \rangle$ is the final average energy of the demon, $T \Delta S$ is the energy lost to the bath during the isothermal process, and $\Delta S$ is the reduction in the demon's entropy. 

Now for the second part of the cycle. The demon lowers its energy levels rapidly to take itself out of equilibrium, so that it can reduce the amount of work it has to do on itself when measuring the system. The crucial question is, what is the maximum work that it can now extract from the target. If the demon lowers all its energy levels to the ground state, then it minimizes the amount of work the measurement must do on itself,  (thus maximizing the net work extraction), and returns to its original configuration, closing the cycle. In doing so the demon reduces its energy to zero, and so retrieves the amount of energy $\langle E \rangle$, being part of the work $W_{\mbox{\scriptsize d}}$. Next, from our analysis above, we know that when it measures the target, the reduction in the entropy of the system can be no more than the resulting increase in the demon's entropy. But recall that in the first part of the cycle, we reduced the demons entropy by $\Delta S$ below the maximal value. During the measurement the demon's entropy cannot therefore increase by more than $\Delta S$, and so the average entropy of the system cannot decrease by more than $\Delta S$. The maximum possible work that the demon can extract from the system is therefore $T \Delta S$. Adding to this the energy that the demon retrieves from itself when it reduces its energy levels, the maximum work extracted by the demon is $W_{\mbox{\scriptsize max}} = \langle E \rangle + T \Delta S = W_{\mbox{\scriptsize d}}$. And this is precisely the minimum amount of work required to raise the energy levels of the demon to begin with. 

We can summarize the above analysis as follows. In order to extract work from a target, the demon must reduce the energy levels of its memory to take itself out of equilibrium. Part of the work that would be required to return these energy levels to their original values is the store of work that the demon possesses, and it is this that is ultimately turned into work. We can see in a more direct way that the demon has a store of work by noting that we can extract $T\Delta S$ from the demon by isothermally expanding it --- this store is measured by the Helmholtz free energy. When we reduce the energy levels of the memory, we reduce its Helmholtz free energy, and this is the work cost of the measurement. We also note that, of the work required to raise the demon's energy levels prior to the measurement, $W_{\mbox{\scriptsize d}} = \langle E \rangle + T \Delta S$, the part that is poured into the bath is the part that is retrieved back from the bath by the demon when it performs the measurement and feedback on the target.  

One can also analyze, in the same way, the work cost that the measurement must pay when it does have access to a cold bath. In this case the amount of work required to raise the demons energy levels to reduce the entropy of the memory by $\Delta S$, is less than $T\Delta S$, and thus the work paid is less. But this makes perfect sense thermodynamically. When a heat engine has access to a cold bath, the amount of heat that it must pour into the hot bath is less than the amount of work it extracts from the cold bath, which is why an engine is useful. The amount of work that must be paid by a measurement to obtain information is precisely the amount of work that a heat engine must pay to extract the same value of work: information remains equivalent to its work value. If the demon has access to a zero temperature bath, then no energy must be paid to extract work, since work is free. 

So how does the energy cost of measurement relate to the usual claim that measurement has no minimum cost, and all the cost is that of erasing of the demon's memory. Erasing the memory means taking it from a high entropy state --- in which it stores all the possible initial states of the target --- to a low entropy state. In our analysis here, the erasure part of the process is when the energy levels of the memory are raised to reduce its entropy, so as to make to the measurement possible. We performed this work at the beginning of the process, but the whole operation is a closed cycle. One could perform it instead at the end, and attribute the cost to erasure. Thus there is a sense in which the attribution of the cost is a matter of interpretation. But there is also a sense in which it is not: the energy cost of the measurement is real, since this is demanded by the first law. Historically, however, people were not concerned with the energy required by the measuring device. The part of the demon's operation that was not understood for so long, was how it was that obtaining a measurement result increased the entropy of the memory. From the point of view of the person who makes the measurement, the memory contains only the measured value, and so appear to have zero entropy. If one analyses the process from this point of view, then it is the erasure step that holds the key to understanding the second law. If this is the central issue, then one tends to consider the target and demon as a unit. When the measurement is made, there need be no net energy change for this unit --- the memory loses work and the target gains it. It is in this sense that the measurement has no cost.  

\appendix*

\vspace{-3mm}

\section{Entropy exchange in a measurement}

\vspace{-3mm}

The initial state of the $M$-dimensional target is $\rho_{\mbox{\scriptsize t}} = \sum_{m=1}^M t_m |m\rangle_{\mbox{\scriptsize t}} \langle m |$, and that of the $N$-dimensional demon is $\rho_{\mbox{\scriptsize d}} = \sum_{n=1}^N d_n |n\rangle_{\mbox{\scriptsize d}} \langle n |$. Thus $|m\rangle_{\mbox{\scriptsize t}}$ are the energy states of the target, and $|n\rangle_{\mbox{\scriptsize d}}$ those of the demon. The initial joint density matrix of the two systems is also diagonal, and these diagonal elements are $\lambda_{mn} = t_m d_n$. The entropy of the joint density matrix is $S_0 =  S_{\mbox{\scriptsize t}} + S_{\mbox{\scriptsize d}}$, with  $S_{\mbox{\scriptsize t}} \equiv S(\rho_{\mbox{\scriptsize t}})$ and $S_{\mbox{\scriptsize d}} \equiv S(\rho_{\mbox{\scriptsize d}})$. To make a measurement on the target, the demon now applies a joint unitary to both systems. After this unitary the systems are correlated, and the demon may apply an independent action on the system for each of a full set of its mutually orthogonal states (since the unitary is arbitrary, we can take these states to be its energy eigenstates without loss of generality). The state of the target for each measurement result, $\rho_n$, is therefore the state within the Hilbert space defined by each energy eigenstate of the demon. We obtain these final target states by projecting the joint density matrix onto the demon's energy eigenstates, and normalizing the result. The probability for obtaining measurement result $n$, $p_n$, is in fact given by this normalization for each $n$~\cite{WMbook}. We will refer to this complete set of projections as a ``projection measurement''. 

Consider the simplest case, in which the joint unitary merely rearranges the eigenvalues of $\sigma^{(0)}$. In this case the measurement is purely classical, and the unitary merely generates classical correlations between the energy bases of the two systems. When we make the projection measurement on the demon, and average over the measurement results, the joint density matrix is left unchanged, and so the final average joint entropy is just the initial joint entropy, $S_0$. The eigenvalues of the density matrix of the target, for each measurement result, are merely a subset of the $\lambda_{mn}$, appropriately normalized. Since the measurement is classical, by rearranging the expression for the entropy of the final average joint density matrix one finds that $S_0 = \langle S_{\mbox{\scriptsize t}}^{\mbox{\scriptsize fin}} \rangle + S_{\mbox{\scriptsize d}}^{\mbox{\scriptsize fin}}$, where $ \langle S_{\mbox{\scriptsize t}}^{\mbox{\scriptsize fin}} \rangle = \sum_n p_n S(\rho_n)$ is the final entropy of the target, averaged over the measurement results, and $S_{\mbox{\scriptsize d}}$ is the entropy of the memory's final state, where in this case the state has been averaged over the measurement results. Since the final state of the memory is pure for each measurement result, this is also the entropy of the probability distribution of the measurement results. Now for the more complex case in which the joint unitary is arbitrary, and changes $\sigma^{(0)}$ to some new state  $\sigma_U$. This new state is no longer diagonal in the energy basis of either system, which in general means that the two systems now have quantum correlations. When we make the projection measurement onto the energy states of the demon, and average over the measurement results, this changes $\sigma_U$ by eliminating various off-diagonal elements. The resulting joint state, $\tilde{\sigma}_U$, is diagonal in the energy basis of the demon, which means it is block-diagonal, where the blocks are the (unnormalized) final density matrices of the target. It is important to note that we can now diagonalize each of these blocks, and this does not change their respective entropies. Once we have done this, $\tilde{\sigma}_U$ is now diagonal, so that we have the relationship $S_{\mbox{\scriptsize joint}} = \langle S_{\mbox{\scriptsize t}}^{\mbox{\scriptsize fin}} \rangle + S_{\mbox{\scriptsize d}}^{\mbox{\scriptsize fin}}$ as before, where now $S_{\mbox{\scriptsize joint}}$ is the entropy of $\tilde{\sigma}_U$. The non-trivial fact that we need now, is the result by Ando that says that averaging a density matrix over a projection measurement never decreases the von Nuemann entropy~\cite{Ando89} (in fact, this is true for all ``bare'' measurements~\cite{Nielsen, FJ}). Thus $S_{\mbox{\scriptsize joint}} \geq S_0$, and so we have $\langle S_{\mbox{\scriptsize t}}^{\mbox{\scriptsize fin}} \rangle + S_{\mbox{\scriptsize d}}^{\mbox{\scriptsize fin}} \; = \; S_{\mbox{\scriptsize joint}} \; \geq \; S_0 \; = \; S_{\mbox{\scriptsize t}} + S_{\mbox{\scriptsize d}}$. Rearranging this gives $\Delta S_{\mbox{\scriptsize d}} \; = \; S_{\mbox{\scriptsize d}}^{\mbox{\scriptsize fin}} - S_{\mbox{\scriptsize d}}  \; \geq \; S_{\mbox{\scriptsize t}} - \langle S_{\mbox{\scriptsize t}}^{\mbox{\scriptsize fin}} \rangle \; = \; - \Delta S_{\mbox{\scriptsize t}} $.

\textit{Acknowledgements:} This work was partially supported by the NSF under Project Nos. PHY-0902906 and  PHY-1005571, and by the ARO MURI grant W911NF-11-1-0268. The author thanks Vlatko Vedral and Kavan Modi for hospitality at Oxford, where some of this work was performed. 

\vspace{-3mm}
 

%

\end{document}